\documentclass[%
 reprint,
 amsmath,amssymb,
 aps,
]{revtex4-1}

\usepackage{graphicx}
\usepackage{dcolumn}
\usepackage{bm}
\usepackage{ulem} 
\usepackage{epstopdf}
\usepackage{xcolor}
\usepackage[caption=false]{subfig}

\begin{document}

\title{Photon echo in ring cavity: pulse area approach}

\author{Sergey A. Moiseev }
\email{s.a.moiseev@kazanqc.org}
\author{Ravil V. Urmancheev }
\email{ravil@kazanqc.org}
\affiliation{
Kazan Quantum Center, Kazan National Research Technical University n.a. A.N.Tupolev-KAI, 10 K. Marx, Kazan 420111, Russia}

\date{\today}

\begin{abstract}
    Pulse area approach has been established as a versatile analytical tool for studying the resonant interaction between the light and the resonant atomic ensemble.
    In recent years photon and spin echoes in cavity assisted schemes become increasingly interesting. 
    In this article we develop the photon echo pulse area approach to describe primary and multi-pulse echo generation in the atomic ensemble placed in the ring cavity. 
    We show that the pulse area approach predicts relative echo magnitudes and whether the system is operating in a single- or a multi-pulse generation regime.
    We also analyze the conditions needed for the realization of these  generation regimes.
    This work develops the pulse area theorem approach for analytical study of  photon/spin echoes in optical and microwave cavities and echo based protocols of quantum memory. 
\end{abstract}

\maketitle

\section{Introduction}
Photon echo \cite{KopvillemPhEcho1963,Kurnit1964} 
is an optical realization of the Hahn spin echo \cite{Hahn1950},
 which is a coherent response of inhomogeneously broadened resonant atomic ensemble to the action of two or more resonant light pulses.
Since its discovery in the beginning of the second half of the previous century  it has been established as a reliable and developing tool in nonlinear coherent spectroscopy, used to measure transition relaxation times and quantum dynamics of different resonant media \cite{Smallwood_2018,cundiff2013optical,Kosarev2019,Beica2020,JIN2020126780,Welinski2020,10.21468/SciPostPhys.12.6.193}.
It has also become the basis of the number of photon echo quantum memory protocols \cite{Moiseev2001,Tittel2010,Ma2021-gn,Horvath2021,Khabat2016,CHANELIERE2018}.
Recent development of modern optical and microwave integral technologies initiates study of photon/spin echoes in optical and microwave cavities \cite{Zhong2017,Heller2020,Moiseev2021,Bustard2022}, which are especially important for elaboration  of quantum memory devices.

Description of the photon echo is based on  a solution of complicated nonlinear Maxwell-Bloch equations, which often compels to use only numerical methods \cite{Weichselbaumer2020}.
In optically dense media, the task is also complicated by the presence of strong rephasing pulses also known as $\pi$-pulses that control atomic coherence.
The pulse area theorem \cite{McCallHahn1969} was proposed to partly bypass these difficulties and provide an analytical tool to study general nonlinear properties of resonant pulse propagation.
The theorem was later developed to consider photon (spin) echoes \cite{Hahn1971,Moiseev1987,Moiseev2020}, three-level systems \cite{Eberly2002} and atoms placed in a Fabry-Perot cavity \cite{Moiseev2022}.

In this work we develop the pulse area theorem to analytically study the echo generation in a single-mode ring cavity, similar to our previous work on Fabry-Perot cavity\cite{Moiseev2022}.
We obtain the general equation describing the area of any pulse during a two-pulse echo generation.
We then solve this equation analytically for the incoming signal pulses and numerically for the first three echo pulses. 
We also provide an approximate analytical solution for the primary echo pulse. 
This allows us to study the conditions for single- and multi-pulse echo generation that are in agreement with experimental investigation \cite{Weichselbaumer2020}.

\section{Pulse area theorem }

\subsection{Basic equations}

We consider an ensemble of N two-level atoms that is placed inside a single-mode ring cavity with the mode central frequency $\omega_0$ being in resonance with the atomic transition.
The atoms  occupy a length $L$ along the optical axis  $z$; $L$ is greater than the light wavelength $\lambda$ and smaller that the cavity length $L_c$.
We assume that the inhomogeneous broadening of the atomic transition $\Delta_{inh}\gg \gamma,$ where $\gamma=1/T_2$ is the homogeneous linewidth and $T_2$ is the coherence time of a single atom, $T_2$ typically shorter than the lifetime of the optical transition  $T_1$.
The electric field is described by a slowly varying amplitude $\mathcal{E}(t)=\mathcal{E}_0 a(t)$, where  
$\mathcal{E}_0=\sqrt{\frac{\hbar\omega_0}{2\varepsilon_0  (\varepsilon L+ L_v)S}}$, $\varepsilon_0$ and $\varepsilon$ are the vacuum and the atomic medium permittivities, $L_v=L_c-L$ and $V=SL_c$ denotes the mode volume.
We assume a uniform excitation of the sample, so the coupling constant  of the dipole interaction between the cavity mode and an atom $g$ is the same for all the atoms: $g = \frac{d}{\hbar}\mathcal{E}_0$, where
$d$ is the dipole moment of the atomic transition.

We use the quantum Tavis-Cummings  model for the interaction of $N$ two-level atoms with the 
cavity mode and apply the input-output formalism of quantum optics \cite{Walls2008} co couple the amplitudes of the cavity mode $\mathcal{E}(t)$ to the amplitudes of the  input $\mathcal{E}_{in}(t)$ and output  $\mathcal{E}_{out}(t)$ field modes (where $\mathcal{E}_{in,out}(t)=\sqrt{\frac{\pi\hbar\omega_0}{\varepsilon_0 S}}a_{in,out}(t)$,  $S$ is a cross-section of the light beam).
In the limit of large number of atoms \cite{Keeling2009}, these transfers to the system of semiclassical Maxwell-Bloch equations \cite{McCallHahn1969,allen1975} for atoms and the resonator field mode:

\begin{align}
\partial_t a & = -\frac{\kappa+\kappa_{in}}{2}a+N g \langle v \rangle_{\Delta} +\sqrt{\kappa}a_{in}, 
\label{eq:field1}
\\
\partial_t u & = - \Delta v - \gamma u,
\\
\partial_t v & = \Delta u - \gamma v + \Omega(t) w,
\\ 
\partial_t w & = -\Omega(t) v,
\\
a_{out} & = \sqrt{\kappa}a - a_{in},
\label{eq:field_in_out}
\end{align}

\noindent
where $\Omega (t)=g a(t)$, $\kappa$ is a decay rate of cavity mode to the external waveguide modes and $\kappa_{in}$ - internal losses of the cavity;
$u,v$ and $w$ are the components of the Bloch vector, dependant on time $t$, and detuning $\Delta$ of the atom; $\langle v \rangle_{\Delta} \equiv \int G(\Delta) v(t, \Delta)  d\Delta ,$ where $G(\Delta)$ is the inhomogeneous line shape. 
Equation (\ref{eq:field_in_out}) relates the cavity mode $a$ to the input and output modes $a_{in}$,  $a_{out}$ according to input-output approach \cite{Walls2008}.

We assume that a pulse of light comes at the moment $t_c = (t_0+t_1)/2$, where $t_0, t_1$ are two distant time moments.
There might have been additional pulses before the time $t_0$ or after $t_1$, but there are no other pulses in the time interval $(t_0,t_1)$, which is much longer that the pulse duration $\delta t$. 
Now we multiply both parts of the field equations \eqref{eq:field1} \&\eqref{eq:field_in_out} by $g$ and integrate over time $\int_{t_0}^{t_1} dt$ to arrive to the general pulse area equation:
    \begin{align}  
        \frac{\kappa_S}{2} \Theta & = \sqrt{\kappa} \Theta_{in} + Ng^2\int_{t_0}^{t_1} dt \int d\Delta  G(\Delta) v(t,  \Delta), 
    \label{eq:field_areas}
    \\
        \Theta_{out} & =\Theta_{in} - \sqrt{\kappa} \Theta,
    \end{align}

\noindent
where $\kappa_S = \kappa+\kappa_{in}$. 
We also considered that $\int_{t_0}^{t_1} \partial_t a(t) dt = a(t_1)-a(t_0)=0$ when $t_1-t_0 \gg \delta t$.

The treatment of this equation is analogous to the case of photon echo area theorem in free space \cite{UrmancheevOptExprs,Moiseev2020} and Fabry-Perot cavity \cite{Moiseev2020} obtained previously. 
This allows us to obtain the general equation for the pulse area of an incoming pulse or an arbitrary echo pulse
(even in the presence of an external exciting pulse with a pulse area $\Theta_{in}$ at the cavity entrance):

    \begin{align}
    \begin{split}
        \frac{\kappa_S}{2} \Theta & = \sqrt{\kappa} \Theta_{in} + \frac{\varkappa}{2}  \Big[  2v_0 \cos^2 \frac{\Theta}{2} +w_0\sin\Theta \Big],
    \\
    \Theta_{out} & =\Theta_{in} - \sqrt{\kappa} \Theta,
    \end{split}
    \label{eq:gen_eq_set_fin}
    \end{align}
where $v_0$ and $w_0$ are resonant components of atomic polarization that phase and lead to the emission of an echo signal at the time $t_c$; $\varkappa=2Ng^2\pi G(0)=2Ng^2/\Delta_{inh}$ for the Lorentzian inhomogeneous broadening $G(\Delta)$. 
Under classical treatment $\varkappa$ corresponds to  atomic absorption per round trip inside the cavity:
$\varkappa=2\alpha L/t_{rt}$ ($\alpha L$ is the optical depth of the resonance transition, $t_{rt}$ is the cavity round trip time).

\subsection{Incoming pulses}

The first incoming pulse arrives to the system in the ground state, so we substitute $v_0 = 0, w_0 = -1$ into Eq. \eqref{eq:gen_eq_set_fin} to get \cite{Chaneliere:14}:

\begin{align}
     \frac{\kappa_{S}}{2} \Theta_1 = \sqrt{\kappa} \Theta_{in,1} - \frac{\varkappa}{2}  \sin\Theta_1 ,
    \label{eq:First_Pulse}
\end{align}

\begin{figure}
\centering
    \includegraphics[width=\linewidth]{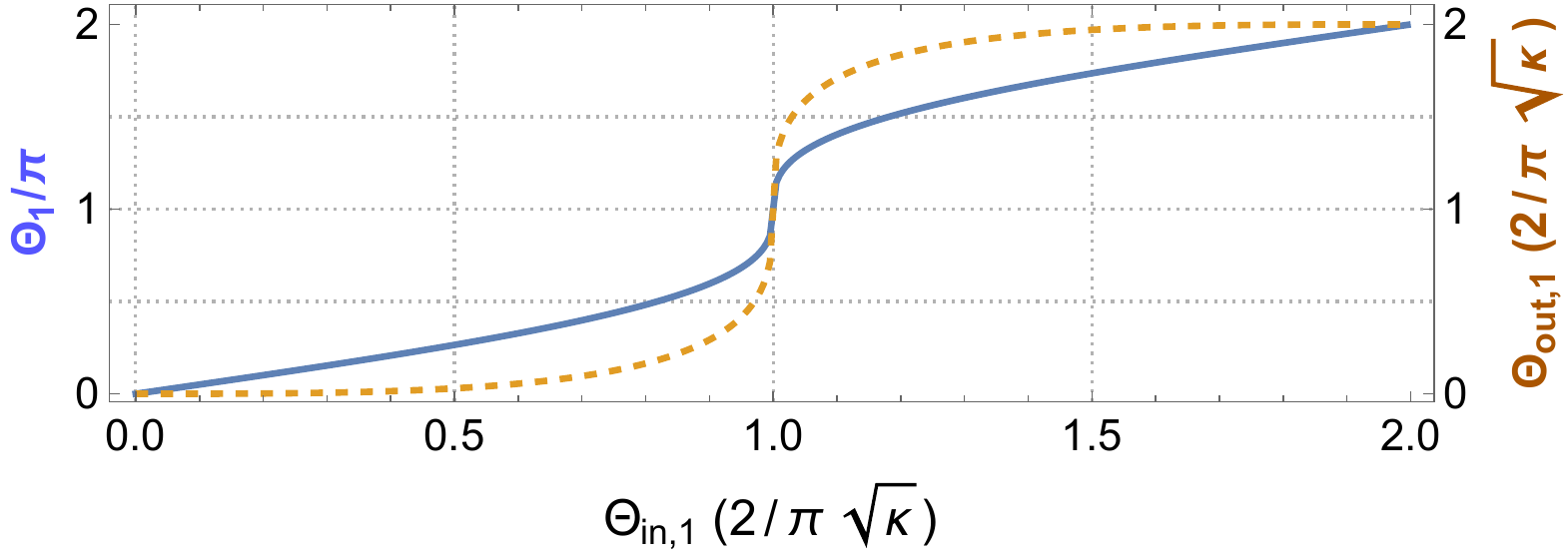}
    \caption{Transition pulse areas $\Theta_1, \Theta_{out,1}$ versus the incoming pulse area $\Theta_{in,1}$ in the impedance-matched ring cavity ($\varkappa = \kappa_S$) in normalized coordinates. Blue solid line shows the pulse area inside the cavity $\Theta_{1}/\pi$ and the orange dashed line shows the pulse area at the output of the cavity $\Theta_{out,1}(2/\pi\sqrt{\kappa})$. The $(2/\sqrt{\kappa})$ factor conforms the areas inside and outside of the cavity.}
\label{fig:Transition_R}
\end{figure}
  
  In the limit of weak signal pulse 
  $\Theta_1\ll1$ we have 
    $\Theta_{out,1} =
  \frac{\varkappa+\kappa_{in}-\kappa}{\kappa_S+ \varkappa} \Theta_{in,1}$.
Analogous to electronics we can derive an impedance matching condition, when $\Theta_{out,1}=0$ and the incoming pulse is fully absorbed: $\xi_{im}=\frac{ \kappa}{\varkappa+\kappa_{in}}=1$.
Parameter  $\xi_{im}$  defines the relative coupling strength of the interaction between the light mode with atoms and free propagating modes, which plays an important role in light-atoms dynamics.
Behavior of pulse area in the impedance matching condition was studied in \cite{Chaneliere:14} for $\kappa_{in}=0$.

Figure \ref{fig:Transition_R} shows the dependence of the transition pulse area inside the ring cavity $\Theta_1$ and at the output of the cavity $\Theta_{out,1}$.
Note that pulse areas outside and inside the cavity have different units and we use the additional factor $(2/\sqrt{\kappa})$ for pulse areas outside the cavity to compensate for that.
The most notable feature of Fig.\ref{fig:Transition_R} is the sharp rise of the output pulse area near the point $(2/\sqrt{\kappa})\Theta_{in,1}=\pi$. 
The sharpness of the rise depends on the ratio between the coupling constants $\xi = \varkappa/\kappa_S$ and is the sharpest at the impedance matching condition.
This feature is a more pronounced version of the self-induced transparency effect in free space \cite{McCallHahn1969}:
if the incoming pulse area is smaller than the threshold value $(2/\sqrt{\kappa})\Theta_{in,1}=\pi$ then the output pulse area gravitates to 0, otherwise it comes close to $2\pi (\sqrt{\kappa}/2)$. 
The introduction of the cavity increases the field-atom interaction and thus the nonlinear effects typical for the optically dense media appear.

For the second pulse we get an equation similar to \eqref{eq:First_Pulse}
\begin{equation}
    \frac{\kappa_S}{2} \Theta_2 = \sqrt{\kappa} \Theta_{in,2} - \frac{\varkappa}{2}  \cos\Theta_{1} \sin\Theta_2,
    \label{eq:Second_Pulse}
\end{equation}
where initial atomic inversion is  modified  by the action of first pulse $w_0=-\cos\Theta_{1}$.
Solution of Eq. \eqref{eq:Second_Pulse}
 is very similar to solving of Eq. \eqref{eq:First_Pulse} when $\Theta_{1}<\pi/2$.
Below we will immediately move on to the study of the behavior of echo signals

\section{Photon echo pulse area}
\subsection{Total echo pulse area}
Using Eqs. \eqref{eq:First_Pulse},\eqref{eq:Second_Pulse} for the incoming pulse we can derive the pulse area of the total pulse area of all echo pulses excited inside the two-level medium. 
To do this we view both incoming pulses as a single composite pulse and write Eq.\eqref{eq:First_Pulse} for this pulse:

\begin{align}
     \frac{\kappa_{S}}{2} \Theta_{tot} = \sqrt{\kappa} (\Theta_{in,1}+\Theta_{in,2}) - \frac{\varkappa}{2}  \sin\Theta_{tot},
    \label{eq:Total_area}
\end{align}
here $\Theta_{tot}$ is the total pulse area of all pulses excited inside the cavity. 
To calculate the total pulse area of all echoes inside and outside the cavity we just need to subtract the areas of the incoming pulses:
\begin{align}
     \Theta_{e,\Sigma}  = & \Theta_{tot}-\Theta_{1}-\Theta_{2},
     \\
     \Theta_{out,\Sigma} = & \Theta_{out,tot}-\Theta_{out,1}-\Theta_{out,2}.
    \label{eq:Tot_echo_area}
\end{align}

This can be used as an estimation tool for the echo pulse area. 
However Eq. \eqref{eq:gen_eq_set_fin} allows to study each echo signal individually.

\subsection{Linear solution}
\label{sec:Approx}

Before studying the particular echo signals let us analyse some general properties of basic Eq. \eqref{eq:gen_eq_set_fin}.
An echo signal is irradiated without presence of external light pulse ($\Theta_{in,e}=0$): 
\begin{equation}
\Theta_e  =  \xi \Big[  2v_0 \cos^2 \frac{\Theta_e}{2} +w_0\sin\Theta_e \Big],
\label{eq:echo_general}
\end{equation}
where $\xi = \varkappa/\kappa_S$.

We can find from \eqref{eq:echo_general} that there is a particular solution:
$\Theta_e=2\pi$ at $v_0\xi=\pi$, that is not longer available in its immediate vicinity and is therefore practically unrealizable. 
The solutions $\Theta_e=\pi,3\pi$ are also impossible, so realistically we have $0\leq\Theta_e<\pi$ and $\pi<\Theta_e<3\pi$.

It is also convenient to rewrite Eq. \eqref{eq:echo_general} in the form
\begin{equation}
\Theta_e  =  2\xi \sqrt{v_0^2+w_0^2} \cos \frac{\Theta_e}{2} \sin\frac{(\varphi-\Theta_e)}{2},
\label{eq:echo_alt}
\end{equation}

\noindent
where $\varphi=2\arctan\{-\frac{v_0}{w_0}\}$ and from Eq. \eqref{eq:echo_alt} we find that $\Theta_e<\varphi$.

Let us first consider the case of a small signal pulse $\Theta_1$, which is typical for optical quantum storage. 
It is particularly interesting since in this regime the echo pulse shape can be  identical to the incoming (first) pulse shape, which means that the pulse area also characterises the energy and efficiency of the outgoing pulse. 
Consider at the beginning the limit of a weak signal pulse $\Theta_1$ and an echo pulse $v_0 \ll 1$ and also that $\Theta_e<1$. 
Then we can obtain the linear solution by keeping only the terms of the first order $O(v_0)$ in  \eqref{eq:echo_general}:

\begin{equation}
  \Theta_e  = \frac{2\xi}{1-\xi w_0}v_0,
  \label{eq:first_order_sol}
\end{equation}
\noindent

Let us consider two  interesting cases. 
The first is the formation of a two-pulse (primary) echo, and the second is the restoration of a suppressed echo (so called ROSE-protocol).
We assume that the  control exciting pulses are launched orthogonally to the resonator axis and have specified pulse areas $\Theta_2, \Theta_3$.

By taking into account the relaxation of the atomic coherence for the time of the primary echo emission  $t_{e1} = 2\tau$, where $\tau$ is the time interval between the pulses, we have $v_{0,pe}=\Gamma_\tau \sin\Theta_1\sin^2\tfrac{\Theta_2}{2}$, where $\Gamma_\tau = e^{-\gamma t_{e1}}$ is the decoherence term and $w_{0,pe} = -\cos\Theta_1\cos\Theta_2$ \cite{Moiseev1987,Moiseev2020}.
By assuming weak pulse area in first signal pulse ($\Theta_1 \ll1 $) for the primary echo we have: $v_{0,pe}\cong \Gamma_{\tau} \Theta_1 \sin^2 (\Theta_2/2)$, $w_{0,pe}=- \cos \Theta_1 \cos \Theta_2\cong-\cos \Theta_2$ 
and for ROSE-echo \cite{Minnegaliev_2021}:
$v_{0,rose}=\Gamma_\tau \sin \Theta_1 \sin^2 (\Theta_2/2)\sin^2 (\Theta_3/2)\cong \Gamma_{2\tau} \Theta_1 \sin^2 (\Theta_2/2)\sin^2 (\Theta_3/2) $, $w_{0,rose}=- \cos \Theta_1 \cos \Theta_2 \cos \Theta_3\cong - \cos \Theta_2 \cos \Theta_3$.
Using these formulas and assuming  $\Theta_2=\Theta_3=\pi$ we get: $v_{0,pe}=v_{0,rose}=\Theta_1$ and $w_{0,pe}=1$, $w_{0,rose}=-1$.
Substituting these values in Eq. \eqref{eq:first_order_sol}  we obtain for the  primary echo: 

\begin{align}
\Theta_{pe}=\frac{2\xi \Gamma_{\tau}}{1-\xi}\Theta_1=
\frac{2\frac{\varkappa}{k}\xi_{im}\Gamma_{\tau}}{ (1+2\frac{k_{in}}{k})\xi_{im}-1}\Theta_1=
\nonumber
\\
=\frac{\varkappa}{k_{in}}\Gamma_{\tau}\Theta_1=
\frac{\varkappa}{\varkappa+k_{in}}
\frac{k}{k_{in}} \Gamma_{\tau}\Theta_1,
\label{Theta_pe}
\end{align}
and for the ROSE echo: 

\begin{align}
\begin{split}
\Theta_{rose}= & \frac{2\xi \Gamma_{2\tau}}{1+\xi}\Theta_1= \frac{\varkappa}{\varkappa+k_{in}}\frac{2 \Gamma_{2\tau}}{1+\xi_{im}}\Theta_1
\\
=& \frac{\varkappa \Gamma_{2\tau}}{\varkappa+k_{in}}\Theta_1,
\end{split}
\label{Theta_rose}
\end{align}
where we have taken into account the impedance matching condition $\xi_{im}=1$.

Thus, as follows from Eq.\eqref{Theta_pe}, the pulse area of the primary echo $\Theta_{pe}$ increases with decreasing resonator losses $k_{in}$ and large values may be larger than the pulse area of the signal pulse $\Theta_1$ when $k\gg k_{in}$, $\varkappa \sim k_{in}$.
This behavior indicates the amplification of the primary echo signal in an inverted medium where it will already be necessary to solve the nonlinear Eq.\eqref{eq:echo_general}.

As it is seen in Eq.\eqref{Theta_rose}, unlike $\Theta_{pe}$, the pulse area of the ROSE signal $\Theta_{rose}$ is always smaller than the pulse area of the signal pulse $\Theta_1$ and only maximally approaches its value with a decrease of resonator losses. 
This behavior corresponds to the fulfillment of the matching condition in the signal field absorption describes the complete recover of the signal pulse in the ROSE signal which is used in the impedance matched photon echo QM schemes \cite{Afzelius2010,Moiseev2021,Minnegaliev_2021}. 
Thus, pulse area approach provides a simple tool for studies of some basic properties of such QM schemes, especially, depended on nonlinear properties of coherent light-atom interactions.

\subsection{Nonlinear approximate solution}

To obtain a more general analytical solution of Eq.\eqref{eq:first_order_sol} we decompose Eq.\eqref{eq:echo_general} in series about the point $\Theta_e = 0, v_0=0$. 
But this time we keep the terms up to the third order of magnitude  ($O(v_0^3)$ or $O(\Theta_e^3)$).
This way we arrive to the cubic equation: 
\begin{equation}
    \Theta_e^3+\frac{3 v_0}{2w_0} \Theta_e^2 +\frac{6 (1-w_0\xi)}{w_0 \xi} \Theta_e - \frac{12 v_0}{w_0} = 0.
    \label{eq:cubic}
\end{equation}
Introducing $\zeta = (1-w_0\xi)/w_0\xi$, we find the discriminant 
\begin{equation}
    \Delta = -108\Big\{ 8\zeta^3 + \frac{v^2_0}{w^2_0}\left[-\tfrac{3}{4}\zeta^2 + 18\zeta +36 \right] + o(v_0^3) \Big\}.
    \label{eq:Discriminant}
\end{equation}

In the quantum memory case (small incoming pulse and strong rephasing pulse) $\Delta<0$, which means that  Eq.\eqref{eq:cubic} has a single real root:
\begin{align}
\begin{split}
    \Theta_e & = -\frac{v_0}{2 w_0}+\sqrt[3]{-\frac{\Delta_1}{2} + \sqrt{\Delta_0}}+\sqrt[3]{-\frac{\Delta_1}{2} - \sqrt{\Delta_0}},
    \\
    \Delta_1 &= -\frac{v_0}{w_0}\left[ 12 + 3\zeta - \frac{1}{4}\left(\frac{v_0}{w_0}\right)^2\right].
\end{split}
\label{eq:Cubic_Root}
\end{align} 
where we introduced $\Delta_0=-\Delta/108$. 
This formula is valid for $\Theta_{in,1} \ll 1$, increasing the signal area will eventually change the sign of $\Delta$, at this point Eq. \eqref{eq:cubic} starts having three real roots. 

\subsection{Primary echo}
\label{sec:Primary_Echo}
Now we take into account that all light pulses excite resonator mode. 
In our previous paper \cite{Moiseev2020} we described the algorithm of finding resonant components of initial polarization and inversion $v_0$ and $w_0$ for each echo pulse. 
These results can be readily applied here.
Substituting the expression for phasing polarization $v_{0,pe}$ and atomic inversion $w_{0,pe}$ for the primary echo signal (see above) in Eq. \eqref{eq:echo_general} we have the nonlinear equation for the primary echo pulse area for the sample placed inside the single mode ring cavity:

    \begin{align}
    \begin{split}
        \Theta_{e1} = 
        \xi \Big[ 2 \Gamma_\tau & \sin\Theta_1\sin^2\tfrac{\Theta_2}{2} \cos^2 \tfrac{\Theta_{e1}}{2}
        \\
        & -\cos\Theta_1\cos\Theta_2 \sin\Theta_{e1} \Big],
    \end{split}
    \label{eq:Ring_PE}
    \\
    \Theta_{out,e1} & = \sqrt{\kappa} \Theta_{e1}.
    \end{align}
\noindent

Note that Eq. \eqref{eq:Ring_PE}  contains the difference of two terms and when these two terms become equal the echo area becomes zero.

Now we can compare the numerical solution of Eq.\eqref{eq:Ring_PE} with approximate solutions obtained in the previous section. 
To test the accuracy of the approximate solution we study the echo pulse area $\Theta_{out,e1}$ while varying the first pulse's incoming area $\Theta_{in,1}$.
The second pulse area we keep constant $(2/\sqrt{\kappa})\Theta_{in,2} = 0.9\pi$. 
The term $2/\sqrt{\kappa}$ accounts for the fact that pulse area inside the cavity $\Theta$ differs from the incoming pulse area $\Theta_{in}$.
To find the values of $\Theta_1, \Theta_2$ used to calculate $v_0$ and $w_0$ we numerically solve Eqs. \eqref{eq:First_Pulse} and \eqref{eq:Second_Pulse}. 

\begin{figure}
    \centering
    \includegraphics[width = \linewidth]{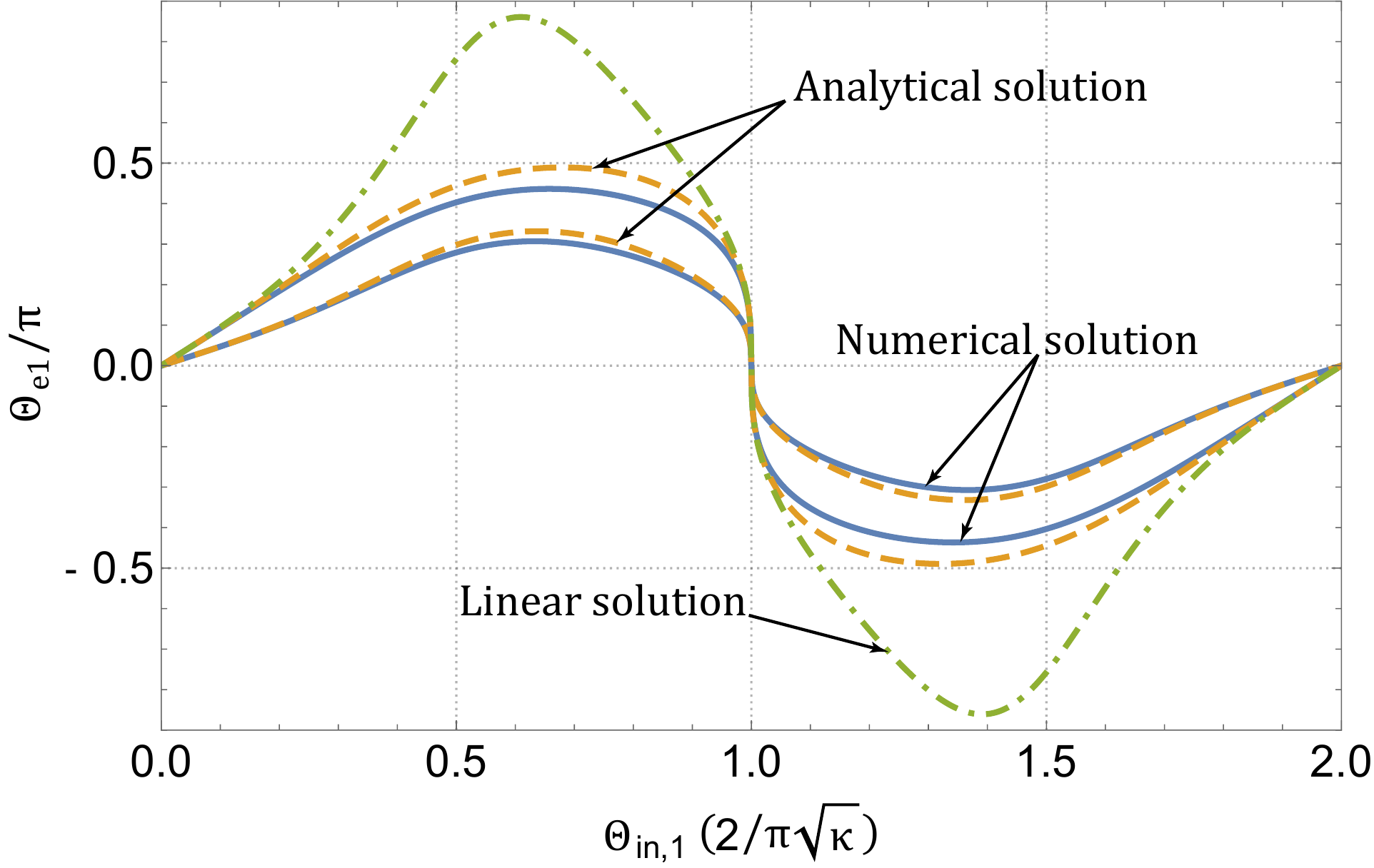}
    \caption{Comparison between the numerical (blue solid curves) solution of Eq.\ref{eq:echo_general}, approximate linear solution Eq.\eqref{eq:first_order_sol} (green dot-dashed line) and approximate cubic solution Eq.\eqref{eq:Cubic_Root} (orange dashed lines); $(2/\sqrt{\kappa})\Theta_{in,2}=0.9\pi, \xi=\varkappa/\kappa=1$.
    The difference between the exact solution and the third order approximation does not exceed 11\%. The error is even less if we introduce the decoherence term $\Gamma_\tau = 0.5$ (the second, closer to the $x-$axis set of solid blue and dashed orange  curves).
    }
    \label{fig:Approx}
\end{figure}

Figure \ref{fig:Approx} shows the comparison between the linear approximation Eq.\eqref{eq:first_order_sol}, cubic approximation Eq.\eqref{eq:Cubic_Root} and numerical solution of Eq.\eqref{eq:Ring_PE}. 
To plot the cubic solution we solved Eq.\eqref{eq:cubic}, which has a single real root for small $\Theta_{in,1}$. 
For larger values of $(2/\sqrt{\kappa})\Theta_{in,1} > 0.75\pi$ Eq. \eqref{eq:cubic} has three real roots from which we choose the root that maintains the continuity of the solution. 
Figure \ref{fig:Approx} shows that the linear approximation works very well in the region $\Theta_{e1}<0.2\pi$, but deviates rapidly after this point. The cubic approximation works well for the primary echo across all range of the incoming first pulse areas $\Theta_1\in[0,2\pi)$, the relative error does not exceed 11\%.
Note that the introduction of decoherence in the form of $\Gamma_\tau=0.5$  term in Eq.\eqref{eq:Ring_PE} brings the cubic solution and numerical solution even closer. 
Decoherence makes the echo signal smaller and consequently the difference between numerical and approximate solutions becomes smaller too (see the caption for the Fig. \ref{fig:Approx}). 

The primary echo experiences a sharp drop and a change of sign near the point $(2/\sqrt{\kappa})\Theta_{in,1}=\pi$. 
This is a consequence of the same nonlinear nature that causes the sharp rise of the transmitted pulse area of a single pulse in Fig. \ref{fig:Transition_R}.
We note that this drop is sharper in the case of the impedance-matched ring cavity than in the case of Fabri-Perot cavity \cite{Moiseev2022}.
The sharpness is higher in the case of the ring cavity due to the uniform excitation of the sample.
In the Fabry-Perot cavity the field forms a standing wave, causing additional phase mismatch between different regions of the sample \cite{Moiseev2022}. 
The peak value of the echo pulse area depends on the coupling constants ratio $\xi=\varkappa/\kappa_S$. 
The echo pulse area is limited and does not exceed $\pi$ for the impedance matched cavity.

\begin{figure}
    \centering
    \includegraphics[width=\linewidth]{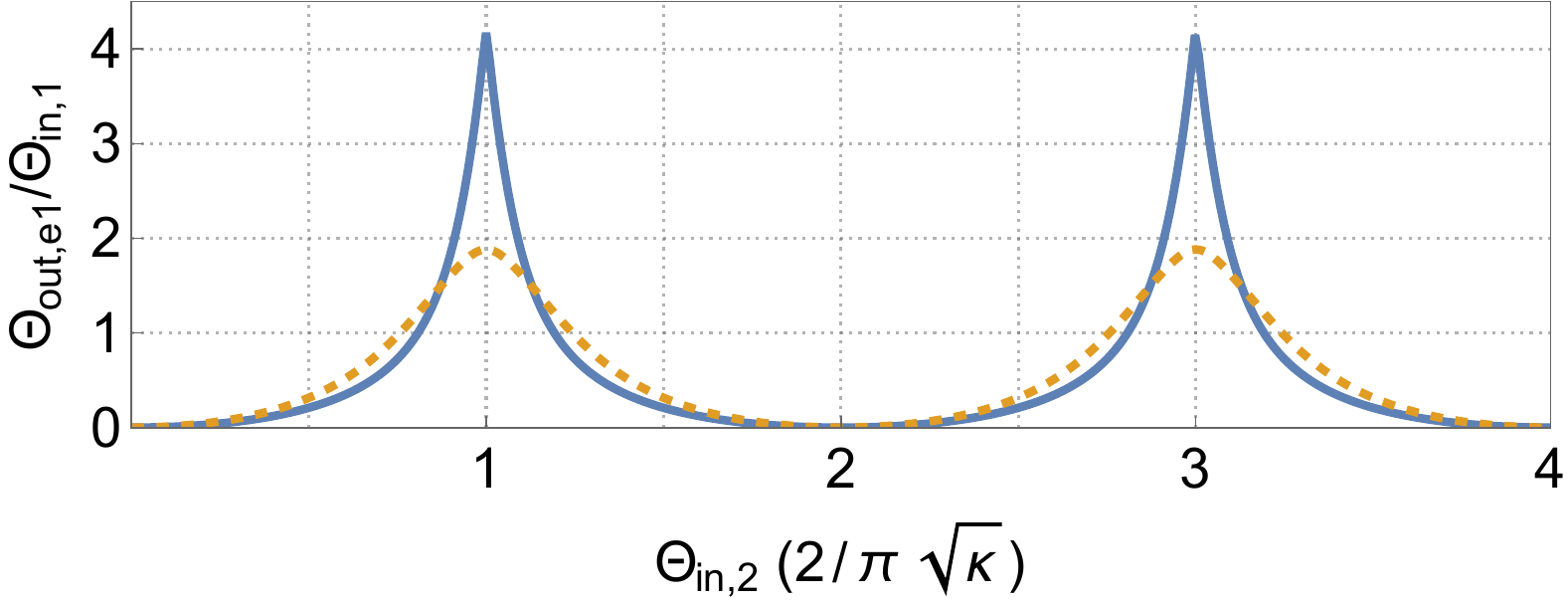}
    \caption{Relative primary photon echo pulse area $\Theta_{out,e1}/\Theta_{in,1}$ at the output of the impedance-matched ring cavity $\varkappa=\kappa_S$ versus the second incoming pulse area $\Theta_{in,2}$ for $(2/\sqrt{\kappa})\Theta_{in,1}=\pi/5$ (blue solid curve) and $(2/\sqrt{\kappa})\Theta_{in,1}=\pi/2$ (orange dashed curve).}
    \label{fig:Echo_Th2}
\end{figure}

Figure \ref{fig:Echo_Th2} shows the dependence of the primary echo area efficiency $\Theta_{out,e1}/\Theta_{in,1}$ versus the incoming pulse area of the second pulse $\Theta_{in,2}$ for the impedance matched cavity $\xi=1$ for negligibly low resonator losses  $k_{in}\ll k$.
The incoming signal pulse area is constant $(2/\sqrt{\kappa})\Theta_{in,1}=\pi/10$ .
Figure \ref{fig:Echo_Th2} shows that the dependence is periodic and peaks near characteristic points $(2/\sqrt{\kappa})\Theta_{in,2}=\pi \cdot n$ where the efficiency exceeds $1$, meaning that the echo pulse is amplified, similar to the free space echoes \cite{Urmancheev2019,Moiseev2020}.
The observed pulse area amplification is also confirmed by the echo pulse energy investigations \cite{1998-Wang-OC,cornish2000demonstration,PhysRevA.79.053851}.
The periodicity is broken in Fabry-Perot cavity, when the field forms a standing wave inside the cavity, leading to a more complicated dependence \cite{Moiseev2022}.

\subsection{Secondary echoes}
After the primary echo pulses there may be secondary echo pulses generated in the medium under certain conditions.
To calculate the pulse area of these echoes we need to find the corresponding phasing resonant components of $v_0$ and $w_0$ at the time of the pulse emission.

For the second echo at the time $3\tau$ we have \cite{Moiseev2020}:
    
    \begin{align}
        \begin{split}
            v_{0,e2} = & \Gamma^2_\tau \cos\Theta_1 \sin\Theta_2 \sin^2\tfrac{\Theta_{e1}}{2} 
            \\
            & + \tfrac{1}{2} \Gamma^2_\tau \sin\Theta_1 \sin\Theta_2 \sin\Theta_{e1},
        \end{split}
        \label{eq:App_Sec_V}
        \\
        \begin{split}
            w_{0,e2} = & -\cos\Theta_1 \cos\Theta_2 \cos\Theta_{e1} 
            \\
            & -\Gamma^2_\tau \sin\Theta_1 \sin^2\tfrac{\Theta_2}{2} \sin\Theta_{e1}.
        \end{split}
        \label{eq:R_v0_Sec_W}
    \end{align}

These expressions we substitute into Eq.\eqref{eq:gen_eq_set_fin} and solve numerically for $\Theta_{e3}$ and $\Theta_{out,e3}$.
Similar but more complicated formulas are derived for the third echo pulse  \cite{Moiseev2020}. Those can be found in the Appendix.

\begin{figure}
    \centering
    \includegraphics[width=\linewidth]{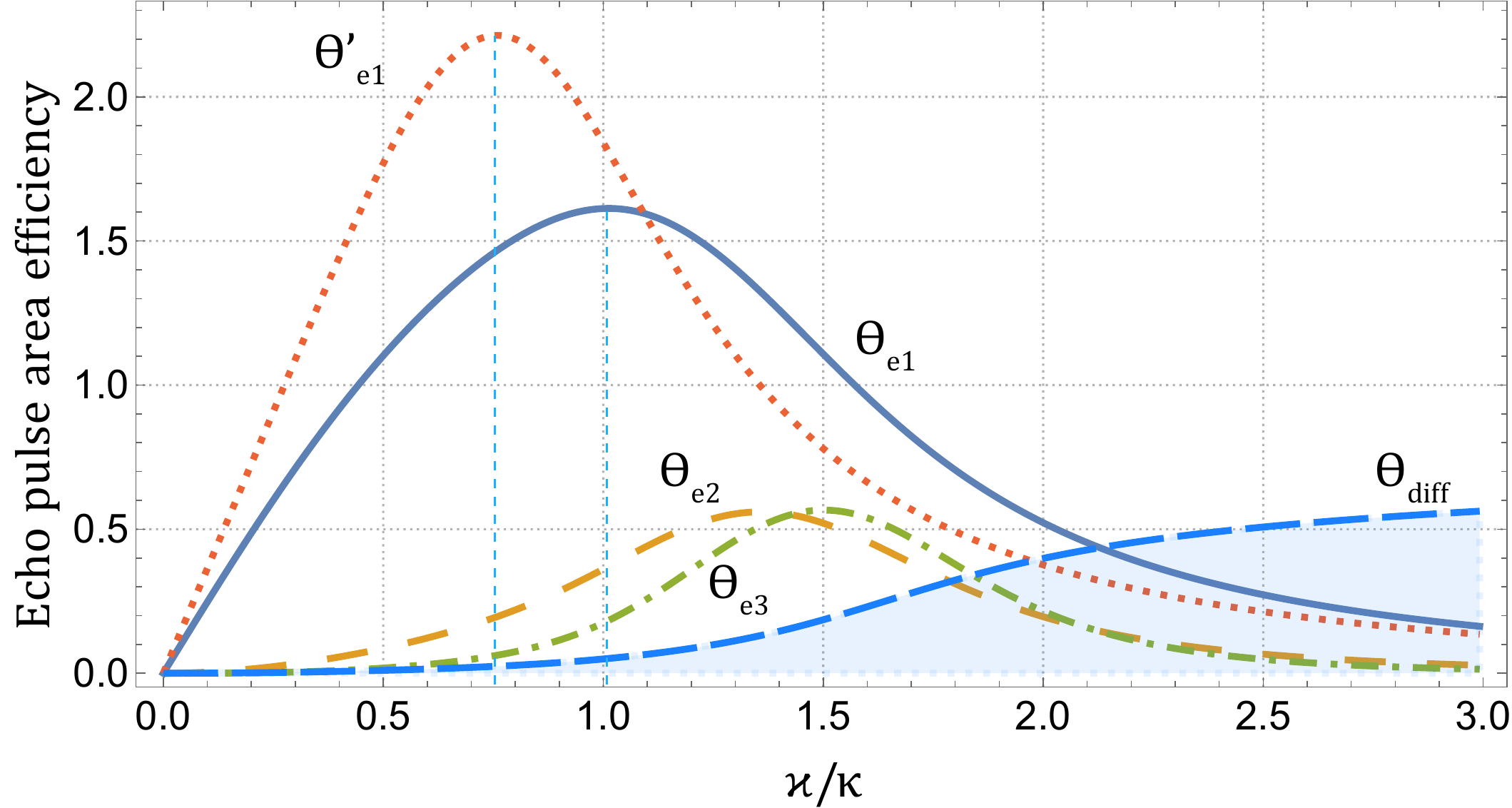}
    \caption{Echo pulse area efficiency $\Theta_{out,e}/\Theta_{in,1}$ for the primary (blue solid line), secondary (orange dashed line), and third (green dash-dotted line)  echoes  versus the coupling strength $\xi =\varkappa/\kappa_S$ in the Ring cavity; $(2/\sqrt{\kappa})\Theta_1=\pi/2, (2/\sqrt{\kappa})\Theta_2=0.9\pi$.
    Red dotted curve marked as $\Theta'_{e1}$ corresponds to primary echo in the case of lower first incoming pulse area  $(2/\sqrt{\kappa})\Theta_{in,1}=\pi/10$, the second pulse incoming area is the same as for all curves.
    Blue vertical lines show the maximums of the primary echo. 
    The filled dashed blue line marked as "$\Theta_{diff}$" plots the $0.1$ of the pulse area of the additional echo pulses excited after the third echo in a classic $(\pi/2,0.9\pi)$ excitation, obtained as $(\Theta_{out,\Sigma}-\Theta_{out,e1}-\Theta_{out,e2}-\Theta_{out,e3})/(10 \Theta_{in,1})$ using Eq.\eqref{eq:Tot_echo_area}. 
    }
    \label{fig:3Echoes_aL}
\end{figure}

Figure \ref{fig:3Echoes_aL} shows the dependence of the first three echo's pulse areas versus the relative coupling strength $\varkappa/\kappa$. 
The incoming pulse areas are $(2/\sqrt{\kappa})\Theta_{in,1}= \pi/2,\, (2/\sqrt{\kappa})\Theta_{in,2}= 0.9\pi$, close to the classic $(\pi/2,\pi)$ sequence of photon/spin echo \cite{Hahn1950,Kurnit1964,allen1975}.
The echo areas dependencies have two characteristic regions. 
In the small coupling regime ($\varkappa/\kappa<0.5$) the primary photon echo is dominating the process and the majority of the energy is emitted into the first echo pulse.
In the strong coupling regime ($\varkappa/\kappa>1$) all three of the considered echo signals become comparable.
The region ($0.5<\varkappa/\kappa<1$) is an intermediate regime, in which the multi-pulse echoes are still small and each consequent echo is smaller than the previous one: $\Theta_{e,n}<\Theta_{e,n+1}$. 

The coupling strength "chooses" whether the system is in the single pulse and multi-pulse echo generation regime. 
This feature has been confirmed experimentally in the recent works \cite{Weichselbaumer2020,Wang_2022}.
In the strong coupling regime there are multiple echo pulses generated inside the cavity. 
Although we calculate only the first three echo signals, we can show that there are actually more secondary echo signals generated by plotting the difference between the total echo pulse area of all echo pulses generated inside the media.
To do so we use Eq.\eqref{eq:Tot_echo_area}, where we have $\Theta_{diff}=0.1(\Theta_{out,\Sigma}-\Theta_{out,e1}-\Theta_{out,e2}-\Theta_{out,e3})/\Theta_{in,1}$.
The term $0.1$ is to agree the scales of different curves in Fig. \ref{fig:3Echoes_aL}.
$\Theta_{diff}$ is negligible in the single pulse regime meaning there are no additional echo signals excited in the medium.
In the strong coupling regime it rises up to $0.5$ meaning that the total pulse area of the consequent echoes is approximately 5 times larger than the signal pulse area $\Theta_{in,1}$, which suggests the existence of multiple additional echo pulses.

Each echo signal in Fig. \ref{fig:3Echoes_aL} has a single maximum, corresponding to the single impedance matching condition.
Previously we derived the impedance matching condition $\varkappa=\kappa$ for the signal pulse but each echo pulse also has its own impedance matching condition depending on the incoming pulse areas.
We show this by comparing the primary echo curves for two different signal pulse areas $(2/\sqrt{\kappa})\Theta_{in,1}=\pi/10$ (red dotted curve) and $(2/\sqrt{\kappa})\Theta_{in,1}=\pi/2$ (blue solid curve).
The curves peak at different coupling strength, meaning that the primary echo pulse impedance matching condition depends on the incoming signal pulse area. 
In turn the second and third echo signals have their own impedance matching conditions and peak at different coupling strengths (orange dashed and green dash-dotted curves in Fig.\ref{fig:3Echoes_aL}).
This allows to optimize the impedance matching condition to a specific generation regime.

\section{Conclusion}
We applied the photon echo area theorem to the two-level atomic ensemble inside an optical ring cavity to find the pulse areas of the incoming pulses and photon echo pulses up to the third echo pulse. 

The approach successfully represents the nonlinear properties of the echo generation process.
The echo strongly depends on the incoming pulse areas as well as on the coupling constant  between the sample and the cavity mode. 
We see that the pulse area theorem successfully captures the main nonlinear features of the photon echo process: primary echo amplification in the inverse medium and multi-pulse echo signal excitation.
The strong point of the approach is that it remains true for arbitrary input pulse areas.
It can also be useful in the analytical study of nonlinear patterns of photon echo in multi-resonator systems \cite{MoiseevJMO2016,Perminov_2018}, which is the subject of subsequent research.

This work constitutes an important step towards establishing photon echo area theorem approach as a general and reliable tool to analyze various photon echo schemes inside optical cavities and an analytical alternative to the computer simulations.

 This research was supported by the Ministry of Science and Higher Education of the Russian Federation (Reg. number NIOKTR 121020400113-1).

\bibliography{Bibliography}
\bibliographystyle{apsrev4-1}

\clearpage
\section{Appendix}
\label{sec:Appendix}
\renewcommand{\theequation}{A\arabic{equation}}
\setcounter{equation}{0}

Signal components of the resonant polarization and inversion of the third echo signal. We can readily use them in Eq.\eqref{eq:echo_general}.
    \begin{align}
        \begin{split}
            v_0 (7\tau/2,z) = & \tfrac{1}{2} \Gamma_\tau^2 \times
            \\
            & \big[\sin\theta_1\sin\theta_2\cos\theta_{e1}\sin\theta_{e2} \\
            & +\cos\theta_1\sin\theta_2\sin\theta_{e1}\sin\theta_{e2} \\
            & +2\cos\theta_1\cos\theta_2\sin\theta_{e1}\sin^2\tfrac{\theta_{e2}}{2}\big]
        \\
            & + \Gamma_\tau^4    \big[\sin\theta_1\cos^2\tfrac{\theta_2}{2}\sin^2\tfrac{\theta_{e1}}{2}\cos^2\tfrac{\theta_{e2}}{2} \\
            & -\sin\theta_1 \sin^2\tfrac{\theta_2}{2}\cos^2\tfrac{\theta_{e1}}{2}\sin^2\tfrac{\theta_{e2}}{2} \big],
       \end{split}
       \label{eq:App_3rd_Echo_V}
       \\
       \begin{split}
               w_0(7\tau/2,z) = & -\cos\theta_1\cos\theta_2\cos\theta_{e1}\cos\theta_{e2} \\
            & - \Gamma_\tau^2 \times [ \sin\theta_1\sin^2\frac{\theta_2}{2}\sin\theta_{e1}\cos\theta_{e2} \\
            & + \cos\theta_1\sin\theta_2\sin^2\frac{\theta_{e1}}{2}\sin\theta_{e2}  \\
            & + \frac{1}{2}\sin\theta_1\sin\theta_2\sin\theta_{e1}\sin\theta_{e2} ] = \\
            & -v_4 \sin\theta_{e2} + w_4 \cos\theta_{e2}.
        \end{split}
        \label{eq:App_3rd_Echo_W}
    \end{align}

\end{document}